\documentclass[aps,preprint]{revtex4}%
\usepackage{amsfonts}
\usepackage{amsmath}
\usepackage{amssymb}
\usepackage[colorlinks,linkcolor=blue,citecolor=blue,urlcolor=blue]{hyperref}
\usepackage{subfigure}
\usepackage{graphicx}
\usepackage{cleveref}
\usepackage{listings}
\usepackage{xcolor}
\usepackage{array}
\usepackage{float}
\usepackage{url}%
\providecommand{\U}[1]{\protect\rule{.1in}{.1in}}
\begin{document}
\title{Scalarized Einstein-Maxwell-scalar Black Holes in a Cavity}
\author{Feiyu Yao}
\email{yaofeiyu@stu.scu.edu.cn}
\affiliation{Center for Theoretical Physics, College of Physics, Sichuan University,
Chengdu, 610064, China}

\begin{abstract}
In this paper, we study the spontaneous scalarization of Reissner-Nordstr{\"o}%
m (RN) black holes enclosed by a cavity in an Einstein-Maxwell-scalar (EMS)
model with non-minimal couplings between the scalar and Maxwell fields. In
this model, scalar-free RN black holes in a cavity may induce scalarized black
holes due to the presence of a tachyonic instability of the scalar field near
the event horizon. We calculate numerically the black hole solutions, and
investigate the domain of existence, perturbative stability against spherical
perturbations and phase structure. The scalarized solutions are always
thermodynamically preferred over RN black holes in a cavity. In addition, a
reentrant phase transition, composed of a zeroth-order phase transition and a
second-order one, occurs for large enough electric charge $Q$.

\end{abstract}
\maketitle
\tableofcontents

\section{Introduction}

The no-hair theorem is important to understand black hole physics, which
initially states that all black hole can be uniquely characterized by only
three externally observable parameters: mass, electric charge, and angular
momentum. Although the no-hair theorem has been proven for the
Einstein-Maxwell field theory, the advent of hairy black hole solutions in the
context of the Einstein-Yang-Mills theory \cite{Volkov:1989fi,Bizon:1990sr}
prompt people to reconsider the no-hair theorem. Later, black holes with
Skyrme hairs \cite{Luckock:1986tr,Droz:1991cx} and black holes with dilaton
hairs \cite{Kanti:1995vq} were also discovered as counter-examples to the
no-hair theorem. For a recent review, see \cite{Herdeiro:2015waa}.

As a way of formation of hairy black holes, spontaneous scalarization is first
studied for neutron stars in scalar-tensor models \cite{Damour:1993hw}. In
this phenomenon of spontaneous scalarization, the non-minimal coupling of the
scalar field to the Ricci curvature can lead to a certain parameter region,
where scalar-free and scalarized neutron star solutions coexist, and the
scalarized one is energetically favoured. Later, the spontaneous scalarization
of black holes is also found in scalar-tensor models
\cite{Cardoso:2013opa,Cardoso:2013fwa}. Recently, the authors of
\cite{Doneva:2017bvd,Silva:2017uqg,Antoniou:2017acq,Doneva:2018rou,Cunha:2019dwb,Herdeiro:2020wei}
studied the spontaneous scalarization in the extended
Scalar-Tensor-Gauss-Bonnet (eSTGB) gravity, where the scalar field is
non-minimally coupled to the Gauss-Bonnet curvature invariant of the
gravitational sector. However, the numerical challenges of studying dynamical
evolution equations in this model compels people to consider other simpler
model. It is at this juncture that the Einstein-Maxwell-scalar (EMS) models
were introduced to gain a deeper insight into spontaneous scalarization
\cite{Herdeiro:2018wub}. In these models, spontaneous scalarization can be
triggered by the strong non-minimal coupling of the scalar field to the
electromagnetic field. In \cite{Herdeiro:2018wub}, a massless and
non-self-interacting scalar field was considered, and an exponential coupling
function was introduced to ensure a tachyonic instability of
Reissner-Nordstr{\"{o}}m (RN) black holes. In \cite{Hod:2020cal}, the
analytical technique is applyied to solve the Klein--Gordon wave equation for
the non-minimally coupled linearized scalar fields in the spacetimes of
near-extremal supporting black holes. Furthermore, spontaneous scalarization
in the EMS models was discussed in context of coupling functions beyond the
exponential coupling \cite{Fernandes:2019rez,Blazquez-Salcedo:2020nhs}, dyons
including magnetic charges \cite{Astefanesei:2019pfq}, axionic-type couplings
\cite{Fernandes:2019kmh}, massive and self-interacting scalar fields
\cite{Zou:2019bpt,Fernandes:2020gay}, horizonless reflecting stars
\cite{Peng:2019cmm}, linear stability of scalarized black holes
\cite{Myung:2018vug,Myung:2019oua,Zou:2020zxq}, higher dimensional scenario
\cite{Astefanesei:2020qxk}, quasinormal modes of scalarized black holes
\cite{Myung:2018jvi,Blazquez-Salcedo:2020jee}, two U(1) fields
\cite{Myung:2020dqt} and quasi-topological electromagnetism
\cite{Myung:2020ctt}. Moreover, the EMS models with a cosmological constant
are considered in \cite{Brihaye:2019gla,Guo:2021zed}. Recently, the effect of
Non-linear electrodynamics corrections on the EMS models is also studied
\cite{Wang:2020ohb}. Analytic approximations were also used to study
spontaneous scalarization of the EMS models
\cite{Konoplya:2019goy,Hod:2020ljo,Hod:2020ius}.

Over the past four decades a preponderance of evidence has accumulated
suggesting a fundamental relationship between gravitation, thermodynamics, and
quantum theory. This evidence is rooted in our understanding of black holes
and their relationship to quantum physics, and developed into the
sub-discipline of black hole thermodynamics. Since the discover of the area
theorem \cite{Hawking:1971tu} and the Hawking radiation
\cite{Hawking:1974rv,Hawking:1974sw}, the analogy between usual thermodynamics
and black hole thermodynamics is confirmed. Furthermore, the four laws of
black hole mechanics were established in \cite{Bardeen:1973gs} and numerous
studies have focused on this subject. Since it is shown that asymptotically
anti-de Sitter (AdS) black holes are thermodynamically stable and the
Hawking-Page phase transition was revealed in Schwarzschild-AdS black holes
\cite{Hawking:1982dh}, thermodynamic properties of various more complicated
black holes have been studied
\cite{Witten:1998zw,Cvetic:1999ne,Chamblin:1999tk,Chamblin:1999hg,Caldarelli:1999xj,Cai:2001dz,Cvetic:2001bk,Nojiri:2001aj}%

As parallel research with that of AdS black holes, studies of black holes in a
cavity have also attracted a lot attentions since York realized that
Schwarzschild black holes can be thermally stable by placing them inside a
spherical cavity, on the wall of which the metric is fixed \cite{York:1986it}.
The phase structure and transitions of black holes in a cavity were shown to
closely resemble that of the AdS counterparts for Schwarzschild black holes
\cite{York:1986it} and RN black holes
\cite{Braden:1990hw,Carlip:2003ne,Lundgren:2006kt}. Recently, it is found that
the resemblance still exist in the extended phase space \cite{Wang:2020hjw}.
And it was discovered that Gauss-Bonnet black holes in a cavity also have
quite similar phase structure and transitions to the AdS counterparts
\cite{Wang:2019urm}. However, it is shown that the phase structure of
Born-Infeld black holes enclosed in a cavity has dissimilarities from that of
Born-Infeld-AdS black holes \cite{Wang:2019kxp,Liang:2019dni}. Moreover, it is
found that there exist significant differences between the thermodynamic
geometry of RN black holes in a cavity and that of RN-AdS black holes
\cite{Wang:2019cax}, and some dissimilarities between the two cases also occur
for validities of the second thermodynamic law and the weak cosmic censorship
\cite{Wang:2020osg}. These findings motivate us to further explore connections
between thermodynamic properties of black holes and their boundary conditions.
For black holes in a cavity, it is shown that there exist the black hole bomb
setup for Kerr black holes \cite{Press:1972zz,Cardoso:2004nk}. Although it is
shown the black hole bomb still exists for charged black holes in a cavity
\cite{Herdeiro:2013pia,Hod:2013fvl,Dias:2018zjg}, the hairy black holes in a
cavity are also found for Einstein-Maxwell gravity coupled to a charged scalar
field \cite{Dias:2018yey}. Therefore, it is interesting to investigated hairy
black holes in a cavity for the EMS model and compare the results with those
of AdS case \cite{Guo:2021zed}.

Nevertheless, the solutions and thermodynamics of hairy black holes in cavity
have rarely been studied in the context of the EMS model. The rest of this
paper is organized as follows. Section \ref{Sec:EMIC} presents the basics of
the EMS model and provides the equations of motion for the solution ansatz of
interest. In section \ref{Sec:SRNBHIC}, we review the scalar-free RN black
holes in cavity and show our main numerical results for scalarized black hole
solutions in a cavity, which include domains of existence, thermodynamic
preference and effective potentials for radial perturbations. In section
\ref{Sec:PS}, we discuss the phase structure and transitions in a canonical
ensemble. We summarize our results with a brief discussion in section
\ref{Sec:DC}. For simplicity, we set $16\pi G=1$ in this paper.

\section{EMS Model in a Cavity}

\label{Sec:EMIC}

The EMS model describes a real scalar field minimally coupled to Einstein's
gravity and non-minimally coupled to Maxwell's electromagnetism. In a cavity,
the EMS model is described by the action%
\begin{equation}
\mathcal{S=}\int_{\mathcal{M}}d^{4}x\sqrt{-g}\left[  R-2\partial_{\mu}%
\phi\partial^{\mu}\phi-f\left(  \phi\right)  F^{\mu\nu}F_{\mu\nu}\right]
+\mathcal{S}_{surf}, \label{eq:IS}%
\end{equation}
where $f\left(  \phi\right)  $ is the coupling function governing the
non-minimal coupling of $\phi$\ and $A_{\mu}$, $F_{\mu\nu}=\partial_{\mu
}A_{\nu}-\partial_{\nu}A_{\mu}$ is the electromagnetic field strength tensor.
And $\mathcal{S}_{surf}$ is the surface terms on $\partial\mathcal{M}$, which
does not affect the equations of motion. Therefore, the equations of motion
that follow from the action $\left(  \ref{eq:IS}\right)  $ are%
\begin{align}
R_{\mu\nu}-\frac{1}{2}Rg_{\mu\nu}  &  =\frac{\mathcal{T}_{\mu\nu}}%
{2},\nonumber\\
\partial_{\mu}\left[  \sqrt{-g}f\left(  \phi\right)  F^{\mu\nu}\right]   &
=0,\label{eq:IEOM}\\
\frac{\partial^{\mu}\left(  \sqrt{-g}\partial_{\mu}\phi\right)  }{\sqrt{-g}}
&  =\frac{\dot{f}\left(  \phi\right)  F^{\mu\nu}F_{\mu\nu}}{4},\nonumber
\end{align}
where $\dot{f}\left(  \phi\right)  \equiv df\left(  \phi\right)  /d\phi$, and
the energy-momentum tensor is%
\begin{equation}
\mathcal{T}_{\mu\nu}=4\left(  \partial_{\mu}\phi\partial_{\nu}\phi
-\frac{g_{\mu\nu}\partial_{\rho}\phi\partial^{\rho}\phi}{2}\right)  +f\left(
\phi\right)  \left[  4F_{\mu\rho}F_{\nu}^{\text{ }\rho}-F^{\mu\nu}F_{\mu\nu
}\right]  .
\end{equation}

The generic spherically symmetric metric can be written as%
\begin{equation}
ds^{2}=-N\left(  r\right)  e^{-2\delta\left(  r\right)  }dt^{2}+\frac{dr^{2}%
}{N\left(  r\right)  }+r^{2}\left(  d\theta^{2}+\sin^{2}\theta d\varphi
^{2}\right)  , \label{eq:metric}%
\end{equation}
The electromagnetic field and the scalar field are given by $A_{\mu}dx^{\mu
}=V\left(  r\right)  dt$ and $\phi=\phi\left(  r\right)  $, respectively. Then
the equations of motion $\left(  \ref{eq:IEOM}\right)  $ reduce to%
\begin{align}
-1+N\left(  r\right)  +rN^{\prime}\left(  r\right)  +r^{2}N\left(  r\right)
\phi^{\prime}\left(  r\right)   &  =-\frac{Q^{2}}{r^{2}f\left(  \phi\right)
},\nonumber\\
\left[  r^{2}N\left(  r\right)  \phi^{\prime}\left(  r\right)  \right]
^{\prime}+r^{3}\phi^{\prime}\left(  r\right)  ^{3}N\left(  r\right)   &
=-\frac{Q^{2}\dot{f}\left(  \phi\right)  }{2f^{2}\left(  \phi\right)  r^{2}%
},\nonumber\\
\delta^{\prime}\left(  r\right)   &  =-r\phi^{\prime}\left(  r\right)
^{2},\label{eq:EoM}\\
V^{\prime}\left(  r\right)   &  =-\frac{e^{-\delta\left(  r\right)  }Q}%
{r^{2}f\left(  \phi\right)  },\nonumber
\end{align}
where primes denote derivatives with respect to the radial coordinate $r$, and
$Q$ is a constant that can be interpreted as the electric charge.

\section{Scalarized RN Black Hole in a Cavity}

\label{Sec:SRNBHIC}

In this paper, we study spontaneous scalarization of the EMS model in a
cavity, which requires a scalar-free solution. When the scalar field $\phi=0$,
the static spherically symmetric RN black hole solution was derived in
\cite{Braden:1990hw},%
\begin{equation}
N\left(  r\right)  =\left(  1-\frac{r_{+}}{r}\right)  \left(  1-\frac
{Q_{b}^{2}}{r_{+}r}\right)  \equiv1-\frac{2m(r)}{r},\text{ }V\left(  r\right)
=-\frac{Q_{b}}{r}dt,\text{ }\delta\left(  r\right)  =0, \label{eq:SFmetric}%
\end{equation}
where $Q_{b}$ is the black hole charge, and $r_{+}$ is the radius of the outer
event horizon. The Hawking temperature $T_{b}$ of the RN black hole is given
by%
\begin{equation}
T_{b}=\frac{1}{4\pi r_{+}}\left(  1-\frac{Q_{b}^{2}}{r_{+}^{2}}\right)  .
\end{equation}
In this scalar-free solution background, one can linearize the scalar equation
in eqn. $\left(  \ref{eq:IEOM}\right)  $ around the scalar-free solution,
which gives%
\begin{equation}
\frac{\partial^{\mu}\left(  \sqrt{-g}\partial_{\mu}\delta\phi\right)  }%
{\sqrt{-g}}=\mu_{eff}^{2}\text{ }, \label{eq:Seqn}%
\end{equation}
where $\mu_{eff}^{2}=-\ddot{f}\left(  0\right)  Q^{2}/\left(  2r^{4}\right)
$. If $\mu_{eff}^{2}<0$, i.e. $\ddot{f}\left(  0\right)  >0$, a tachyonic
instability is induced, and a scalarized black hole solution can bifurcate
from the scalar-free RN black hole solution. In the remainder of the paper, we
focus an exponential coupling, $f\left(  \phi\right)  =e^{\alpha\phi^{2}}$
with $\alpha>0$, which satisfies $\dot{f}\left(  0\right)  =0$ and $\ddot
{f}\left(  0\right)  >0$.

To obtain non-trivial hairy black hole solutions of the non-linear ordinary
differential equations $\left(  \ref{eq:EoM}\right)  $, one needs to impose
regular boundary conditions at the event horizon and the boundary of the
cavity. The regularity of the solutions across the event horizon at $r=r_{+}$
gives that the solutions can be approximated by a power series expansion in
$r-r_{+}$,%
\begin{align}
m\left(  r\right)   &  =\frac{r_{+}}{2}+\left(  r-r_{+}\right)  m_{1}%
+...\text{, }\delta\left(  r\right)  =\delta_{0}+\left(  r-r_{+}\right)
\delta_{1}+...,\nonumber\\
\phi\left(  r\right)   &  =\phi_{0}+\left(  r-r_{+}\right)  \phi
_{1}+...\text{, }V\left(  r\right)  =\left(  r-r_{+}\right)  v_{1}+...\text{,
} \label{eq:SolHor}%
\end{align}
where%
\begin{equation}
m_{1}=\frac{Q^{2}}{2r_{+}^{2}f\left(  \phi_{0}\right)  },\text{ }\phi
_{1}=-\frac{Q^{2}\dot{f}\left(  \phi_{0}\right)  }{2\left[  f^{2}\left(
\phi_{0}\right)  r_{+}^{3}-f\left(  \phi_{0}\right)  r_{+}Q^{2}\right]
},\text{ }\delta_{1}=-r_{+}\phi_{1}^{2},\text{ }v_{1}=\frac{Qe^{-\delta_{0}}%
}{r_{+}^{2}f\left(  \phi_{0}\right)  }.
\end{equation}
The parameters $\phi_{0}$ and $\delta_{0}$ determine the expansion
coefficients and hence the solutions in the vicinity of the horizon. Outside
the cavity, we require that the scalar field vanishes and it recover the
scalar-free solution \cite{Dias:2018yey}. Therefore the boundary conditions at
the boundary of the cavity is%
\[
\phi\left(  r_{B}\right)  =0,\delta\left(  r_{B}\right)  =0.
\]
We can use a standard shooting method to solve eqn. $\left(  \ref{eq:EoM}%
\right)  $ for a family of black hole solutions with the boundary conditions
at the event horizon and the boundary of the cavity. Note that the solutions
and the associated physical quantities scale as%
\begin{equation}
r\rightarrow\lambda r,\phi\rightarrow\phi,m\rightarrow\lambda m,V\rightarrow
V,\delta\rightarrow\delta,Q\rightarrow\lambda Q,M\rightarrow\lambda M,
\label{eq:scale}%
\end{equation}
where $\lambda$ is a constant and the ADM mass $M$ is introduced by the
asymptotic expansion of the solutions at spatial infinity,%
\begin{equation}
m\left(  r\right)  =M-\frac{Q^{2}+Q_{s}^{2}}{2r}+...,
\end{equation}
.So we introduce some reduced quantities for later use,%
\begin{equation}
q=\frac{Q}{M},a_{H}=\frac{A_{H}}{16\pi M^{2}},t_{H}=8\pi MT_{H},
\end{equation}
which are dimensionless and invariant under the scaling symmetry.

To study how a scalarized black hole solution bifurcates from a scalar-free
solution, we derive the zero modes of the scalar perturbation in the
scalar-free black hole background. First, we express the scalar perturbation
as a spherical harmonics decomposition%
\begin{equation}
\delta\phi=%
%TCIMACRO{\dsum \limits_{l,m}}%
%BeginExpansion
{\displaystyle\sum\limits_{l,m}}
%EndExpansion
Y_{lm}\left(  \theta,\phi\right)  U_{l}\left(  r\right)  .
\end{equation}
With this decomposition, the scalar equation $\left(  \ref{eq:Seqn}\right)  $
then simplifies to%
\begin{equation}
\frac{\partial_{r}\left[  r^{2}N\left(  r\right)  U_{l}^{\prime}\left(
r\right)  \right]  }{r^{2}}-\left[  \frac{l\left(  l+1\right)  }{r^{2}}%
+\mu_{eff}^{2}\right]  U_{l}\left(  r\right)  =0,
\end{equation}
where $N\left(  r\right)  $ is given by eqn. $\left(  \ref{eq:SFmetric}%
\right)  $. For the black hole in a cavity, $U_{l}\left(  r\right)  $ is
regular at $r=r_{+}$ and vanishes at $r=r_{B}$. The boundary conditions of
$U_{l}\left(  r\right)  $ would pick up a set of black hole solution with
different reduced charge $q$ if one fixes $\alpha$ and $l$. The black hole
solutions can be labeled by a non-negative integer $n$, and $n=0$ is the
fundamental mode, whereas $n>0$ corresponds to overtones. In this paper, we
focus on the $l=0=n$ mode since it gives the smallest $q$ of the black hole
solutions for a given $\alpha$ \cite{Herdeiro:2018wub}. The reduced charge
$q_{exist}\left(  \alpha\right)  $ of the $l=0=n$ mode compose the bifurcation
line in the $\alpha-q$ plane, on which scalarized black hole solutions emerge
from the RN black holes in a cavity.\begin{figure}[tb]
\begin{center}
\includegraphics[width=0.48\textwidth]{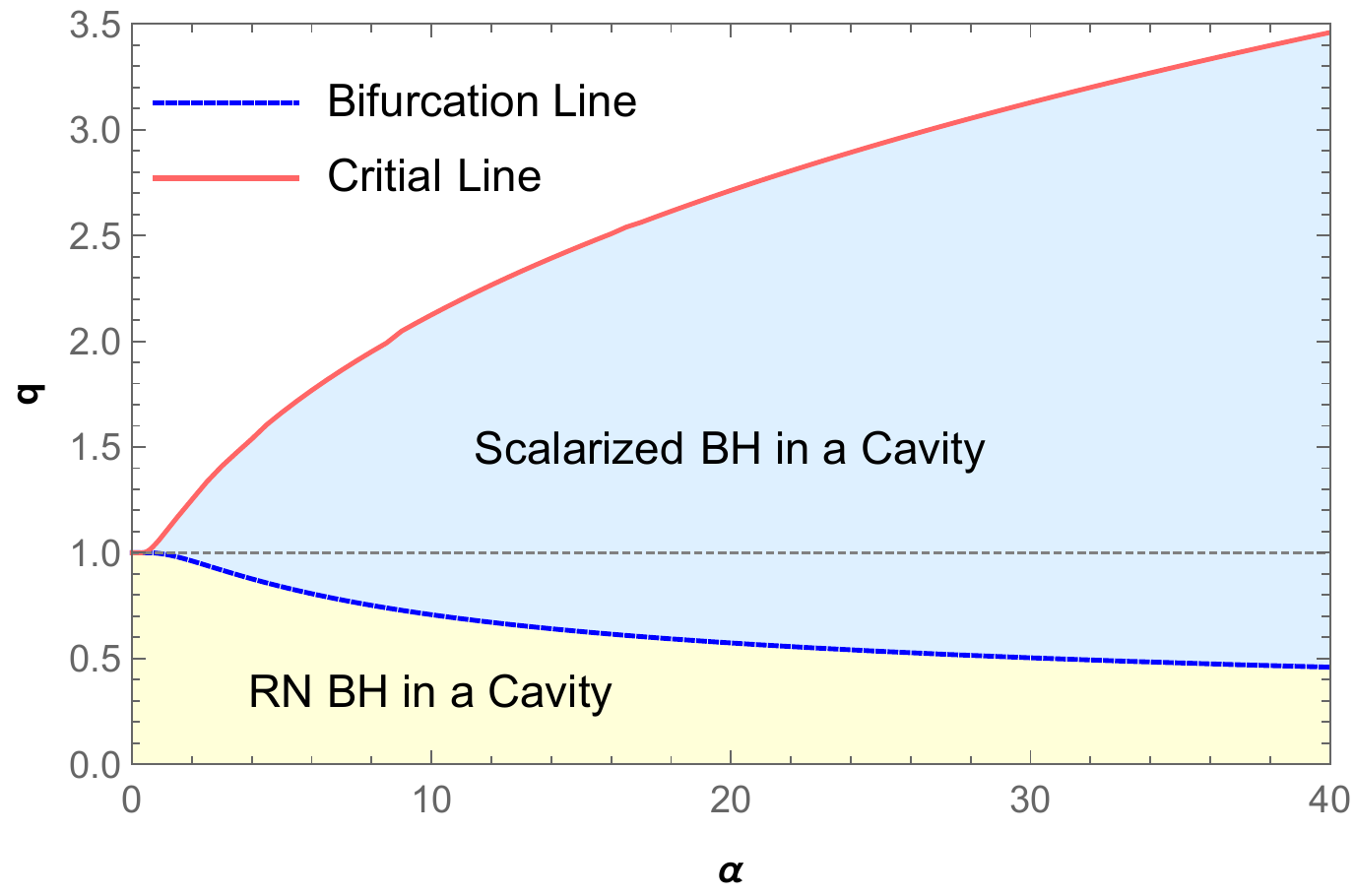}
\includegraphics[width=0.48\textwidth]{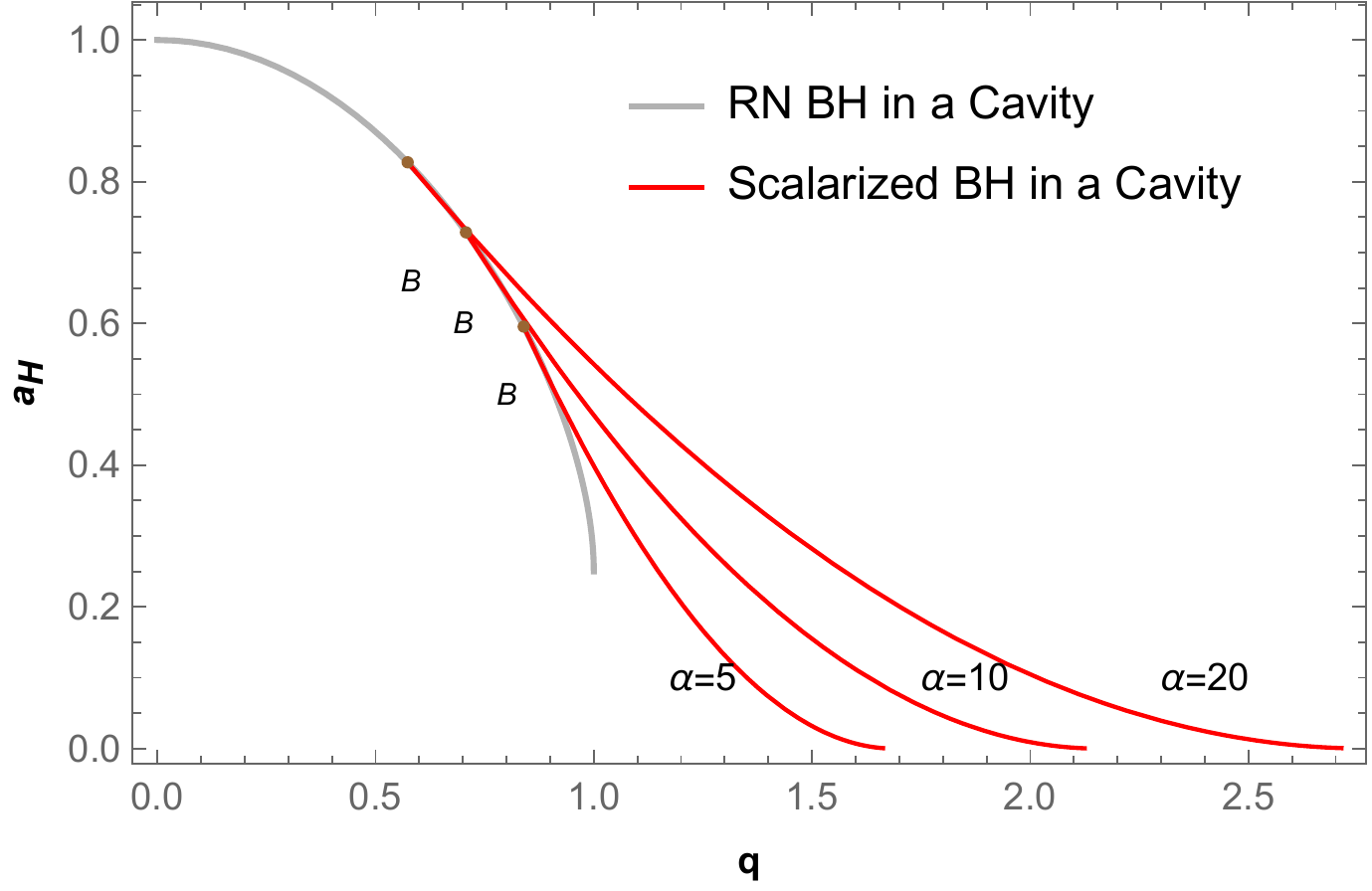}
\end{center}
\caption{{\protect\footnotesize Domain of existence and thermodynamic
preference for scalarized RN BH in a cavity. \textbf{Left panel: }Domain of
existence in the }${\protect\footnotesize \alpha-q}$
{\protect\footnotesize plane, which is displayed by a shaded light blue region
and bounded by the bifurcation and critical lines. The blue dashed line
represents the bifurcation line, where scalarized black holes bifurcate from
RN black holes in a cavity as zero modes. The red line marks critical
configurations of scalarized black holes, where the horizon area vanishes with
the mass remaining finite. The horizontal dashed gray line denotes extremal RN
black holes in cavity, above which RN black hole solutions in a cavity do not
exist. \textbf{Right panel: }Reduced area }${\protect\footnotesize a}%
_{{\protect\footnotesize H}}$ {\protect\footnotesize versus reduced charge
}${\protect\footnotesize q}$ {\protect\footnotesize for RN black holes (a gray
line) and scalarized RN black holes with various values of }%
${\protect\footnotesize q}$ {\protect\footnotesize (red lines). For a given
}${\protect\footnotesize q}${\protect\footnotesize , }%
${\protect\footnotesize a}_{{\protect\footnotesize H}}$
{\protect\footnotesize of the scalarized RN black hole in a cavity is larger
than that of the RN black hole, and increases with the growth of
}${\protect\footnotesize \alpha}${\protect\footnotesize . The scalarized black
hole solutions are always entropically preferred.}}%
\label{fig: Criticalline}%
\end{figure}We present the numerical results, e.g., domains of existence,
thermodynamic preference and effective potentials, for scalarized black hole
solutions in a cavity. We express non-linear differential equation $\left(
\ref{eq:EoM}\right)  $ in terms of a new coordinate%
\begin{equation}
x=1-\frac{r_{+}}{r}\text{ with }0\leq x<1-\frac{r_{+}}{r_{B}},
\end{equation}
and employ the NDSolve function in Wolfram Mathematic to numerically solve the
equations in the interval $10^{-8}<x<1-r_{+}/r_{B}$. In what follows, we
confine ourselves to the simplest case of nodeless, spherically symmetric
black hole solutions and leave general configurations for future
work.\begin{figure}[tb]
\begin{center}
\includegraphics[width=0.65\textwidth]{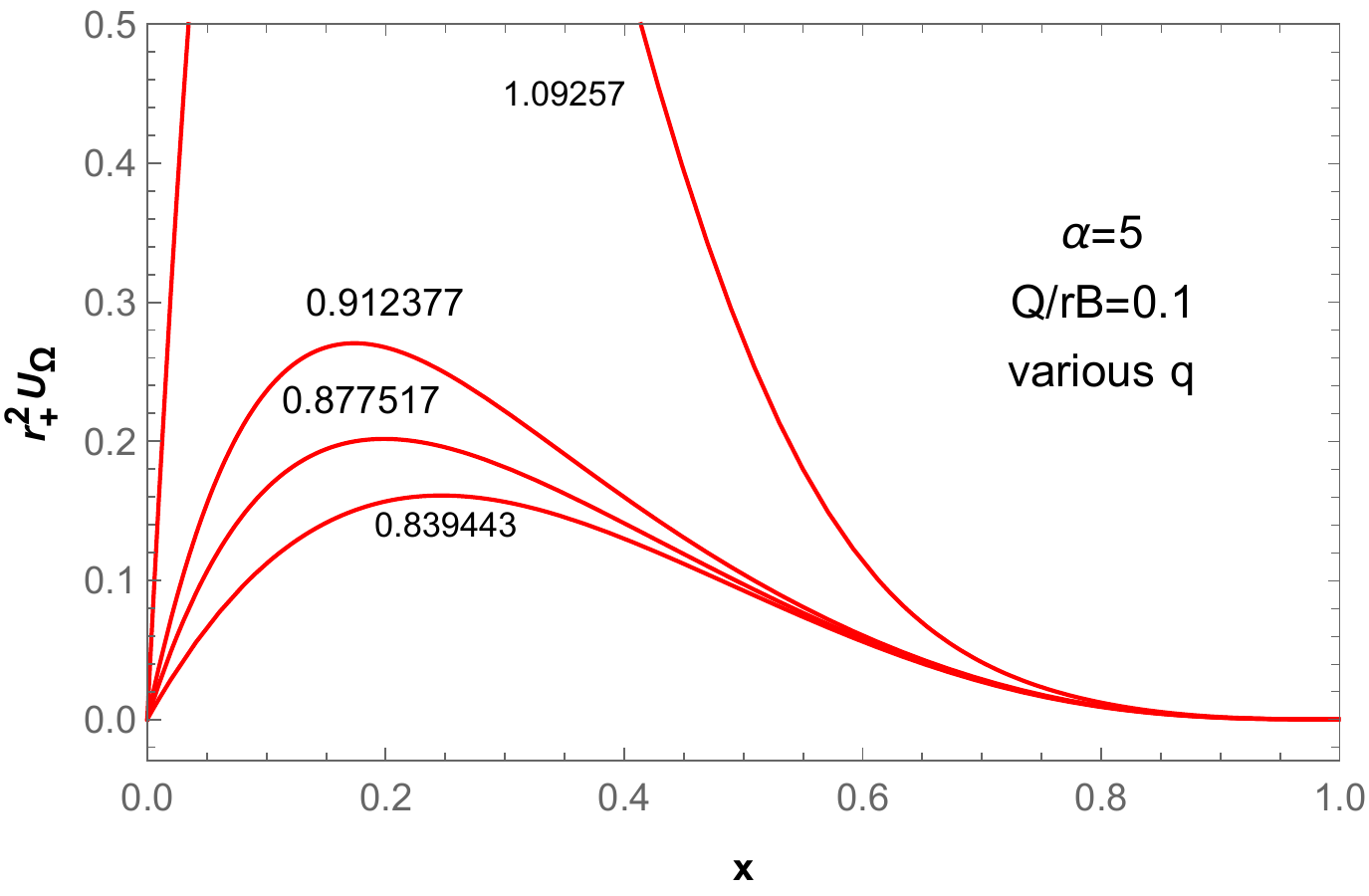}
\end{center}
\caption{{\protect\footnotesize Effective potentials for scalarized RN black
holes with }${\protect\footnotesize \alpha=5}$ {\protect\footnotesize for
various values of }${\protect\footnotesize q}${\protect\footnotesize . They
are all positive definite, which means that the scalarized solutions are
stable against radial perturbations. }}%
\label{fig:effUvarq}%
\end{figure}

In the left panel of FIG. $\left(  \ref{fig: Criticalline}\right)  $, we
present the domain of existence for scalarized RN black holes in a cavity with
$r_{B}/Q=10$. For a given $\alpha$, scalarized solutions emerge from the
bifurcation line as zero modes, and can be continuously induced by increasing
$q$ until they reach the critical line. The numerical results suggest that,
for scalarized solutions on the critical line, the horizon radius $r_{+}$
vanishes, whereas the mass $M$ and the charge $Q$ remain finite. The domain of
existence for scalarized solutions is bounded by the bifurcation and critical
lines, and the boundary contour of domain of existence shows a close
resemblance to that of RN black holes \cite{Herdeiro:2018wub} and RN-AdS black
holes \cite{Guo:2021zed}. The numerical results show that there exists a
unique set of nodeless scalarized solutions for given $\alpha$ and $q$ in the
domain of existence. We plot the reduced area $a_{H}$ of the scalarized
solutions against the reduced charge $q$ for several $\alpha$ values of in the
right panel of FIG. $\left(  \ref{fig: Criticalline}\right)  $, which
indicates that scalarized solutions are entropically preferred over RN black
hole solutions in a cavity.

Then we consider spherically symmetric and time-dependent linear perturbations
around the black hole. The metric ansatz including the perturbations can be
written as \cite{Fernandes:2019rez}%
\begin{equation}
ds^{2}=-\tilde{N}\left(  r,t\right)  e^{\tilde{\delta}\left(  r,t\right)
}dt^{2}+\frac{dr^{2}}{\tilde{N}\left(  r,t\right)  }+r^{2}\left(  d\theta
^{2}+\sin^{2}\theta d\varphi^{2}\right)  , \label{eq:metricp}%
\end{equation}
where%
\begin{equation}
\tilde{N}\left(  r,t\right)  =N\left(  r\right)  +\epsilon\tilde{N}_{1}\left(
r\right)  e^{-i\Omega t}\text{ and }\tilde{\delta}\left(  r,t\right)
=\delta\left(  r\right)  +\epsilon\tilde{\delta}_{1}\left(  r\right)
e^{-i\Omega t}. \label{eq:mNp}%
\end{equation}
The time dependence of the perturbations is assumed to be Fourier modes with
frequency. Similarly, the ansatzes of the scalar and electromagnetic fields
are given by%
\begin{equation}
\tilde{\phi}\left(  r,t\right)  =\phi\left(  r\right)  +\epsilon\tilde{\phi
}_{1}\left(  r\right)  e^{-i\Omega t}\text{ and }\tilde{V}\left(  r,t\right)
=V\left(  r\right)  +\epsilon\tilde{V}_{1}\left(  r\right)  e^{-i\Omega t},
\label{eq:mVp}%
\end{equation}
respectively. Solving eqn. $\left(  \ref{eq:IEOM}\right)  $ with the ansatzes
$\left(  \ref{eq:mNp}\right)  $ and $\left(  \ref{eq:mVp}\right)  $, one can
extract a Schrodinger-like equation for the perturbative scalar field
$\tilde{\phi}_{1}\left(  r\right)  $,%
\begin{equation}
-\frac{d^{2}\Psi\left(  r\right)  }{dr^{\ast2}}+U_{\Omega}\Psi\left(
r\right)  =\Omega^{2}\Psi\left(  r\right)  ,
\end{equation}
where $\Psi\left(  r\right)  \equiv r\phi_{1}\left(  r\right)  $, and the
tortoise coordinates $r^{\ast}$ is defined by $dr^{\ast}/dr=e^{\delta\left(
r\right)  }N^{-1}\left(  r\right)  $. The effective potential $U_{\Omega}$ is
given by%
\[
U_{\Omega}=\frac{e^{-2\delta}N}{r^{2}}\left[  1-N-2r^{2}\phi^{\prime2}%
-\frac{Q^{2}}{r^{2}f\left(  \phi\right)  }\left(  1-2r^{2}\phi^{\prime2}%
+\frac{\ddot{f}\left(  \phi\right)  }{2f\left(  \phi\right)  }+2r\phi^{\prime
}\frac{\dot{f}\left(  \phi\right)  }{f\left(  \phi\right)  }-\frac{\dot{f}%
^{2}\left(  \phi\right)  }{f^{2}\left(  \phi\right)  }\right)  \right]  ,
\]
which can be shown to vanish at the event horizon and spatial infinity. A
positive definite $U_{\Omega}$ ensures that scalarized black hole solutions
are stable against the spherically symmetric perturbations. We display
effective potentials for scalarized solutions with $\alpha=5$ in FIG. $\left(
\ref{fig:effUvarq}\right)  $. The numerics show that the scalarized solutions
always have positive effective potentials, thus are stable against the
spherically symmetric perturbations.

\section{Phase structure}

\label{Sec:PS}

In this section, we consider phase structure and transitions of scalarized and
RN black holes in a cavity. For black holes in a cavity, in a canonical
ensemble, the wall of the cavity, which is located at $r=r_{B}$, is maintained
at a temperature of $T$ and a charge of $Q$. It was showed
\cite{Braden:1990hw} that the system temperature $T$ and charge $Q$ can be
related to the black hole temperature $T_{b}$ and charge $Q_{b}$ as%
\begin{equation}
Q=Q_{b}\text{ and }T=\frac{T_{b}}{\sqrt{N\left(  r_{B}\right)  }},
\end{equation}
respectively. The Helmholtz free energy $F$ and the thermal energy $E$ were
also given in \cite{Carlip:2003ne},%
\begin{equation}
F=r_{B}\sqrt{1-N\left(  r_{B}\right)  }-T\pi r_{+}^{2},\text{ }E=r_{B}%
\sqrt{1-N\left(  r_{B}\right)  }.
\end{equation}
The phase that has the lowest possible Helmholtz free energy $F$ is the
globally stable phase of a multiphase system. The rich phase structure of
black holes comes from solving $T\left(  r_{+}\right)  $ for $r_{+}$. If
$T\left(  r_{+}\right)  $ is a monotonic function with respect to $r_{+}$,
there is only one branch for the black hole. More often, there exists a local
minimum/maximum for $T\left(  r_{+}\right)  $ at $r_{+}=r_{+,\min}/r_{+,\max}%
$. In this case, there is more than one branch for the black hole,
corresponding to different phases. The scalar-free RN black holes in a cavity
has been studied in \cite{Lundgren:2006kt} in a canonical ensemble, it show
that the thermodynamics and phase structure of RN Black Holes in a cavity is
similar to that of AdS counterparts and the critical values is%
\begin{equation}
\tilde{r}_{+c}=5-2\sqrt{5},Q_{c}=\sqrt{5}-2,\tilde{T}_{c}=\frac{\sqrt
{5+2\sqrt{5}}\left(  9-4\sqrt{5}\right)  }{2\left(  5-2\sqrt{5}\right)
^{3}\pi}.
\end{equation}

\begin{figure}[ptb]
\begin{center}
\includegraphics[width=0.96\textwidth]{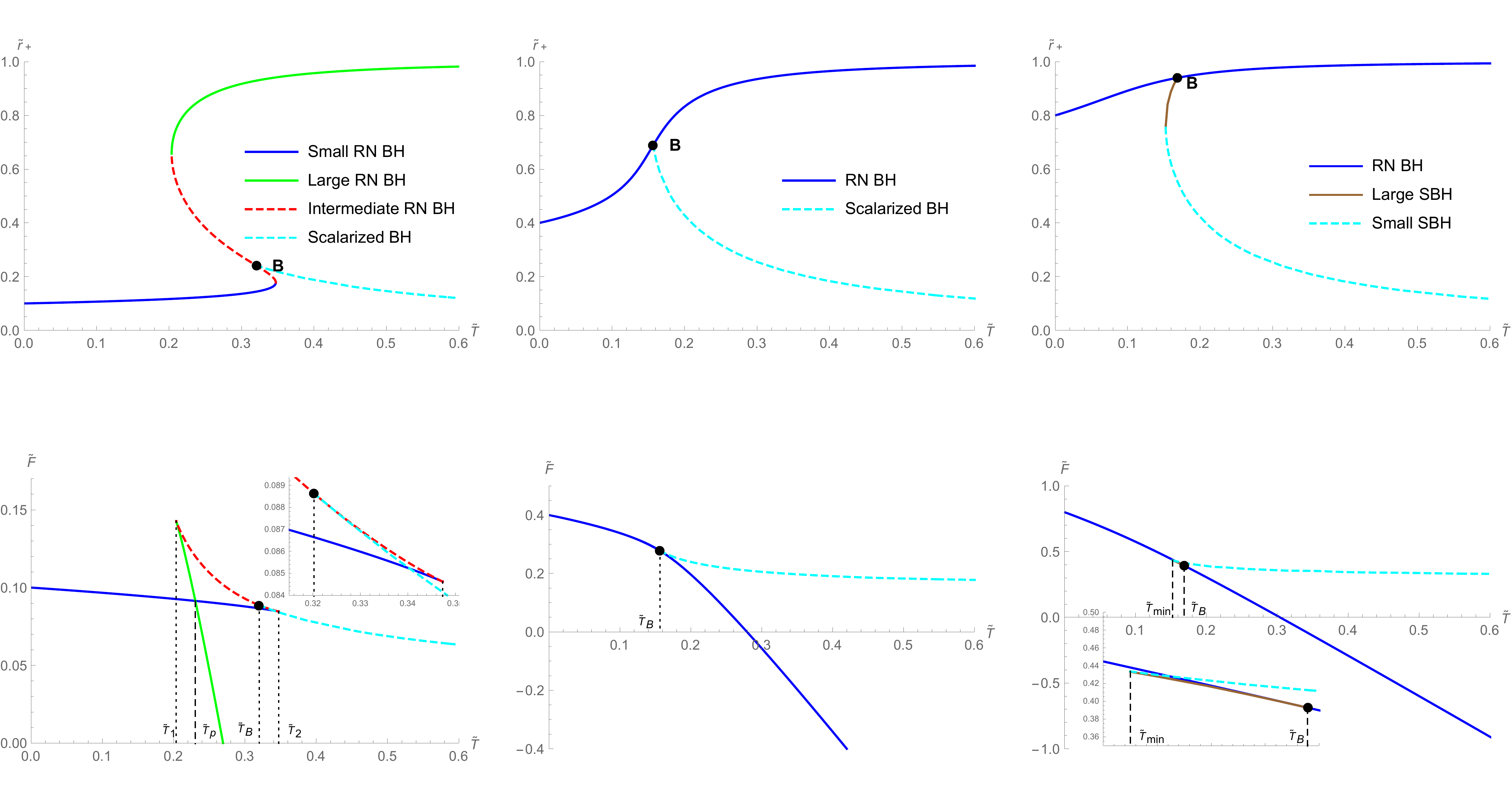}
\end{center}
\caption{{\protect\footnotesize Plots of the reduced horizon radius
}${\protect\footnotesize \tilde{r}}_{+}$ {\protect\footnotesize and the
reduced free energy }${\protect\footnotesize \tilde{F}}$
{\protect\footnotesize against the reduced temperature }%
${\protect\footnotesize \tilde{T}}$ {\protect\footnotesize for RN black holes
and scalarized black holes in a cavity with different values of }%
${\protect\footnotesize \tilde{Q}}$ {\protect\footnotesize and }%
${\protect\footnotesize \alpha=5}${\protect\footnotesize . The top row
displays }${\protect\footnotesize \tilde{r}}_{+}-{\protect\footnotesize \tilde
{T}}$ {\protect\footnotesize for various }${\protect\footnotesize \tilde{Q}}$,
{\protect\footnotesize and the corresponding }${\protect\footnotesize \tilde
{F}}-{\protect\footnotesize \tilde{T}}$ {\protect\footnotesize is shown in the
bottom row. Moreover, the phases plotted by dashed line are thermodynamic
unstable since their heat capacity is negative. And black points
}${\protect\footnotesize B}$ {\protect\footnotesize are}
{\protect\footnotesize bifurcation points. \textbf{Left Column: }%
}${\protect\footnotesize \tilde{Q}=0.1}${\protect\footnotesize . three
scalar-free phases, namely Small RN BH, Intermediate RN BH and Large RN BH,
coexist for }${\protect\footnotesize \tilde{T}}_{1}%
{\protect\footnotesize <\tilde{T}<\tilde{T}}_{2}$
{\protect\footnotesize since} ${\protect\footnotesize \tilde{Q}<\tilde{Q}}%
_{c}${\protect\footnotesize . And the scalarized BH exist when }%
${\protect\footnotesize \tilde{T}>\tilde{T}}_{B}${\protect\footnotesize ,
where }${\protect\footnotesize \tilde{T}}_{B}$ {\protect\footnotesize is the
temperature of bifurcation point.} {\protect\footnotesize And the Small RN BH
and Large RN BH are} {\protect\footnotesize thermodynamic stable, whereas the
Intermediate RN BH and Scalarized BH are} {\protect\footnotesize thermodynamic
unstable. Therefore, there is only a first-order phase transition form Small
RN BH to Large RN BH at }${\protect\footnotesize \tilde{T}=\tilde{T}}_{p}%
${\protect\footnotesize . \textbf{Center Column: }}%
${\protect\footnotesize \tilde{Q}=0.4}$.{\protect\footnotesize \textbf{ }There
are one branch of scalar-free solutions and one branch of scalarized
solutions, dubbed RN BH and Scalarized BH. Since the Scalarized BH is
thermodynamic unstable, there is no phase transition in this case.
\textbf{Right Column: }}${\protect\footnotesize \tilde{Q}=0.8}$.
{\protect\footnotesize There are one branch of scalar-free solutions and two
branch of scalarized solutions, dubbed RN BH, Small Scalarized BH (Small SBH)
and Large Scalarized BH (Large SBH). Note that the Large SBH is thermodynamic
stable. As }${\protect\footnotesize \tilde{T}}$
{\protect\footnotesize increases from }${\protect\footnotesize 0}%
${\protect\footnotesize , the black hole jumps from the RN BH branch to the
Large SBH one, corresponding to the zeroth order phase. Further increasing
}${\protect\footnotesize \tilde{T}}${\protect\footnotesize , there would be a
first order phase transition returning to the RN BH. Therefore there is a RN
BH}${\protect\footnotesize \rightarrow}${\protect\footnotesize Large
SBH}${\protect\footnotesize \rightarrow}${\protect\footnotesize RN BH
reentrant phase transition.}}%
\label{fig:rFTfixQ}%
\end{figure}

Considering the scalarized black hole solution, there is additional branch for
$T\left(  r_{+}\right)  $. Therefore we displays $\tilde{r}_{+}$ and
$\tilde{F}$ against $\tilde{T}$ for scalarized and scalar-free RN black holes
in a cavity with different values of $\tilde{Q}$ and $\alpha=5$ in FIG.
$\left(  \ref{fig:rFTfixQ}\right)  $, where%
\begin{equation}
\tilde{r}_{+}=\frac{r_{+}}{r_{B}},\text{ }\tilde{F}=\frac{F}{r_{B}},\text{
}\tilde{T}=r_{B}T.
\end{equation}
Moreover, the thermodynamic stabilities against thermal fluctuations of these
phase is also shown in FIG. $\left(  \ref{fig:rFTfixQ}\right)  $, where
thermodynamic unstable and stable phase is plotted with dashed and solid line,
respectively. In a canonical ensemble, the quantity we consider is the
specific heat at constant electric charge:%
\begin{equation}
C_{Q}=T\left(  \frac{\partial S}{\partial T}\right)  _{Q}=2\pi r_{+}T\left(
\frac{\partial r_{+}}{\partial T}\right)  _{Q}.
\end{equation}
Since the entropy is proportional to the size of the black hole, a positive
specific heat means that the black hole radiates less when it is smaller.
Thus, the thermal stability of the branch follows from $C_{Q}>0$%
.

When $\tilde{Q}<\tilde{Q}_{c}$, three phases of the scalar-free RN black
holes, namely Small BH, Intermediate BH and Large BH, coexist for some range
of $\tilde{T}$. And there is only one phase of scalarized RN black holes,
which bifurcates form the scalar-free solution. For $\tilde{Q}=0.1<\tilde
{Q}_{c}$, we plot $\tilde{r}_{+}$ and $\tilde{F}$ against $\tilde{T}$ in the
left column of FIG. $\left(  \ref{fig:rFTfixQ}\right)  $, where different
colored lines represent different phases. The three phases of scalar-free
black holes form the characteristic swallowtail, which usually means a van der
Waals-like phase transitions. And the scalarized BH bifurcates form
Intermediate BH. Moreover, the free energy of scalarized BH is always not the
lowest, and the negative specific heat of scalarized BH means that it is
thermodynamic unstable. So the scalarized BH is always not globally stable and
there is only a first-order transition form Small BH to Large BH at $\tilde
{T}=\tilde{T}_{p}$ for $\tilde{Q}<\tilde{Q}_{c}$.

When $\tilde{Q}_{c}<\tilde{Q}<0.49$, there are one branch of scalar-free
solutions and one branch of scalarized solutions. The scalar-free one is
thermodynamic stable and the scalarized one is thermodynamic unstable.
Therefore, the globally stable phase is always scalar-free RN BH, and there is
no phase transition, which is shown in the middle column of FIG. $\left(
\ref{fig:rFTfixQ}\right)  $.\begin{figure}[h]
\begin{center}
\includegraphics[width=0.48\textwidth]{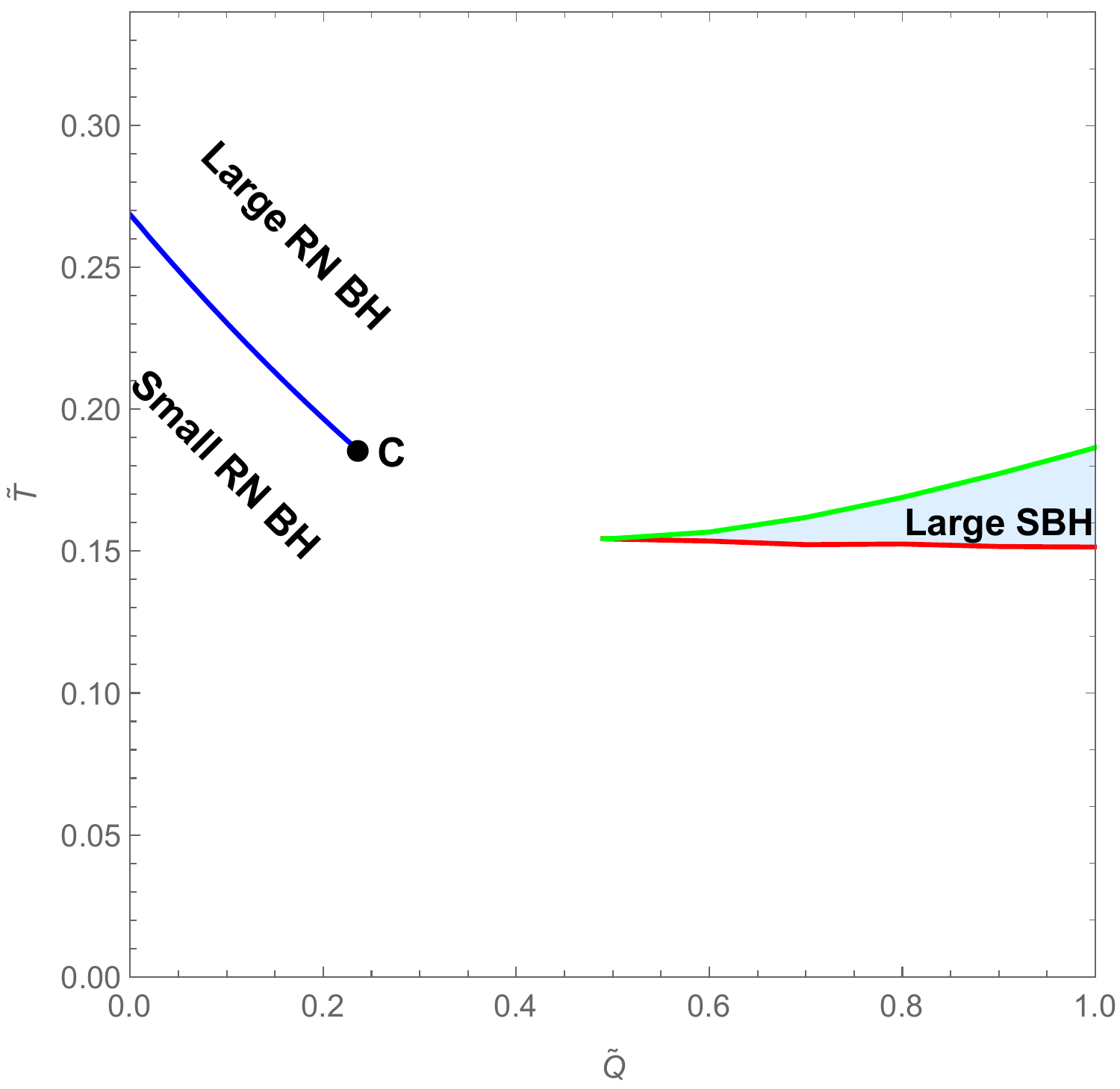}
\end{center}
\caption{{\protect\footnotesize Phase diagrams of RN black holes and
scalarized black holes enclosed by a cavity in the $\tilde{Q}$-$\tilde{T}$
plane with }${\protect\small \alpha=10}${\protect\footnotesize . The phase
diagram exhibits the globally stable phases, which have the lowest free
energy. The first-order phase transition line separating Large BH and Small BH
is displayed by blue lines and terminate at the critical point, marked by
black dots }$C${\protect\footnotesize . In the blue region, the Large
scalarized BH (Large SBH) is globally stable. It shown that a reentrant phase
transition happens when $\tilde{Q}$ is large enough. This reentrant phase
transition consists of a zero-order phase transition and a second-order phase
transition, corresponding the red line and green line, respectively.}}%
\label{fig:PhaseTQ}%
\end{figure}

When $\tilde{Q}>0.49$, there are branches of scalarized black holes, namely
Small SBH and Large SBH, coexisting between $\tilde{T}=\tilde{T}_{\min}$ and
$\tilde{T}=\tilde{T}_{B}$. The Large SBH has positive specific heat and is
thermodynamic stable, whereas the Small SBH is thermodynamic unstable. As
$\tilde{T}$ increase from $0$, there is a jumps from the scalar-free RN BH
branch to the Large SBH one, corresponding to the zeroth order phase
transition between RN BH and Small SBH, followed by a second-order phase
transition returning to RN BH. This RN BH$\rightarrow$Large SBH$\rightarrow$RN
BH phase transition corresponds to a reentrant phase transition.

Furthermore, we display the globally stable phase of black holes in a cavity
in the $\tilde{Q}-\tilde{T}$ plane. In FIG. $\left(  \ref{fig:PhaseTQ}\right)
$, the blue line represents Large BH/Small BH first-order transition lines,
which terminates at the critical point, where a second-order phase transition
occurs. The red line and green line correspond the zeroth order phase
transition from RN BH to Small SBH and the first order phase transition
returning to RN BH, respectively. Comparing with the phase diagram of
scalar-free RN black hole in a cavity, the join of scalarized BH results in a
additional structure for large enough $\tilde{Q}$, corresponding a reentrant
phase transition, which consists of the red line and green line.

\section{Discussion and Conclusion}

\label{Sec:DC}

In this paper, we investigated spontaneous scalarization of charged black
holes enclosed by a cavity in the EMS model with the nonminimal coupling
function $f\left(  \phi\right)  =e^{\alpha\phi^{2}}$, and studied the phase
structure of these black holes in a canonical ensemble. The scalarized black
holes in a cavity can bifurcate from scalar-free solution on the bifurcation
line, which consists of zero modes of the scalar-free solutions. The domains
of existence, thermodynamic preference, radial stability and temperature of
these solutions were numerically investigated for the black holes in a cavity.
In the $\alpha-q$, FIG. $\left(  \ref{fig: Criticalline}\right)  $ shows that
the domain of existence for scalarized RN black holes in a cavity is bounded
by the bifurcation and critical lines, which is very similar with the case of
not in a cavity \cite{Herdeiro:2018wub} and RN-AdS \cite{Guo:2021zed}. Note
that the scalarized solutions in a cavity always have positive effective
potentials, which means they are always stable against the spherically
symmetric perturbations, unlike the case of RN-AdS black holes.

Furthermore, we studied the thermodynamic of these black holes and a richer
phase structure than the scalar-free case is displays in FIG. $\left(
\ref{fig:PhaseTQ}\right)  $. In the small $\tilde{Q}$ regime, scalarized black
holes never globally minimize the free energy, there is a classical
first-order phase transition line separating Large BH and Small BH, resembling
that of the liquid/gas system closely. However, when $\tilde{Q}$ is large
enough, there is another structure since scalarized black holes can be the
globally stable phase in the large $\tilde{Q}$ regime, where the system
undergoes a RN BH$\rightarrow$Large SBH$\rightarrow$RN BH reentrant phase
transition as $\tilde{T}$ increases from $0$. And this reentrant phase
transition consists of a zeroth-order phase transition and a second-order one.

In this paper, the spontaneous scalarization of black holes in a cavity were
studied, and the phase structure is investigated in the normal phase space,
where the cavity radius is fixed. The results closely resemble those of the
AdS counterparts for scalarized RN black holes. Recently, the phase space of
black holes in a cavity has been extended by including a thermodynamic
pressure and a thermodynamic volume \cite{Wang:2020hjw}, it is interesting to
consider scalarized black holes enclosed by a cavity in the extended phase
space. Moreover, one can study excited scalarized solutions since only the
fundamental state was considered in this paper. On the other hand, other
thermodynamic properties could be investigated for scalarized black holes in a
cavity and it would be very interesting to explore these phenomena in the
context of other non-linear electrodynamics black holes in a cavity and check
whether analogies with the AdS counterparts can go beyond RN black holes.

\begin{acknowledgements}
We are grateful to Peng Wang, Guangzhou Guo and Qingyu Gan for useful discussions and valuable comments. This work is supported in part by NSFC (Grant No. 11875196, 11375121, 11947225 and 11005016).
\end{acknowledgements}

\bibliographystyle{unsrturl}
\bibliography{ref}

\end{document}